\def\be{\begin{equation}}
\def\ee{\end{equation}}
\def\bea{\begin{eqnarray}}
\def\eea{\end{eqnarray}}

\documentclass[aps,prl,twocolumn]{revtex4}

\usepackage{epsfig,graphicx}

\begin{document}

\title{Searching for squeezed particle-antiparticle 
correlations in high energy heavy ion collisions}

\author{Sandra ~S. ~Padula$^1$ and  O.~Socolowski, Jr.$^2$}
\affiliation{$^1$Instituto de F\'\i sica Te\'orica-UNESP, C. P. 70532-2, 01156-970 S\~ao Paulo, SP, Brazil\\
$^2$IMEF - FURG - C. P. 474, 96201-900, Rio Grande, RS, Brazil}

\date{\today}

\begin{abstract}
Squeezed correlations of particle-antiparticle pairs were predicted to exist if the hadron masses were modified in the hot and dense medium formed in high energy heavy ion collisions. Although well-established theoretically, they have not yet been observed experimentally.  We suggest here a clear method to search for such signal, by analyzing the squeezed correlation functions in terms of measurable quantities. We illustrate this suggestion for simulated $\phi\phi$ pairs at RHIC energies. 
\end{abstract}


\maketitle

\centerline{\sl \small DOI: 10.1103/PhysRevC.82.034908} 

\section{Introduction}

In the late 90's, it was shown \cite{acg99} that the hadronic mass-modification in hot and dense media 
 could lead to a  a novel type of correlation between bosons and their antiparticles. 
These  {\sl squeezed back-to-back correlations} (BBC) are the result of a quantum mechanical  
transformation relating in-medium quasi-particles to two-mode squeezed states of their free, observable counterparts \cite{acg99}. This is achieved by means of a Bogolioubov-Valatin (BV) transformation linking the asymptotic creation (annihilation) operators, $\hat{a}^\dagger_\mathbf k$ ($\hat{a}_\mathbf k$),  of the observed bosons with momentum $k^\mu\!=\!(\omega_k,{\bf k})$, to the operators, $\hat{b}^\dagger_\mathbf k$ ($\hat{b}_\mathbf k$), corresponding to thermalized quasi-particles in the medium. 
This transformation is given by
$\hat{a}_k=\hat{b}_k c_k + \hat{b}^\dagger_{-k} s^*_{-k} \; ; \;   
\hat{a}^\dagger_k=\hat{b}^\dagger_k c^*_k + \hat{b}_{-k} s_{-k}  \; , \; 
$ 
being $c_k=\cosh(f_k)$ and $s_k=\sinh(f_k)$.  For conciseness, we keep here the short-hand notation introduced in Ref.\cite{acg99},  
where ($-k$)  denotes an opposite sign in the spacial components of the momenta. 
Since the BV transformation between these operators is equivalent to a squeezing operation, the coefficient of this transformation, $ f_{k_i,k_j}(x)=\frac{1}{2}\log\left[\frac{\omega_{k_i}(x)+\omega_{k_j}(x)} {\Omega_{k_i}(x)+\Omega_{k_j}(x)}\right] $ is called squeezing parameter; $\omega_{k_i}^2=m^2+{\mathbf k_i}^2$ and $\Omega_{k_i}^2=m_*^2+{\mathbf k_i}^2$ are, respectively the dispersion relation in terms of the asymptotic mass, $m$,  and in terms of the in-medium modified mass, $m_*$. 
A complete description of the phenomenon should include a parameterization for $m_*$  depending on the particles' 
momenta and on their coordinates in the hot and dense system. For proposing the means to search for it experimentally, however, 
it suffices to asume a linear relation between the masses, i.e., $m_* = m\pm \delta m$,  as was also 
considered in \cite{acg99,pkchp01,phkpc05,pscn08}.

After the publication in Ref. \cite{acg99}, a similar BBC between fermion-antifermion pairs was demonstrated to exist \cite{pkchp01}, 
if the masses of these particles were modified in-medium. Both the fermionic (fBBC)  and the bosonic (bBBC)  back-to-back squeezed correlations are described by analogous formalisms, being both positive correlations with {\sl unlimited} intensity. 
This behavior is in contrast to what is observed in femtoscopy, or Hanbury-Brown and Twiss effect (HBT), a quantum statistical correlations among identical particles. In the HBT realm, 
two bosons with similar momenta are positively correlated, with intensity ranging from 1 to 2, whereas two fermions with similar momenta are anti-correlated, with intensity between 0 and 1. Besides, the fBBC and the bBBC constitute a direct probe of hadronic mass shift in the hot and dense media, contradicting naive expectations that 
such information would vanish at the freeze-out surface, leaving no trace on correlations. 

In the remainder of this paper, we will focus on the bosonic case only, more especifically, on bosons that are their own antiparticles, such as $\phi \phi$ or  $\pi^0 \pi^0$.  
The two-particle correlation function is written as 
\be C_2({\mathbf k_1},{\mathbf k_2})= \frac{N_2(\mathbf k_1,\mathbf k_2)}{N_1(\mathbf k_1) N_1(\mathbf k_2)}, \label{corrgen}
\ee 
where the numerator is the two-particle joint distribution and the denominator is the product of the two single-inclusive distributions.  

The numerator in Eq. (\ref{corrgen}) is proportional to the expectation value of the four-operator, 
$ \langle \hat{a}^\dagger_{\mathbf k_1} \hat{a}^\dagger_{\mathbf k_2} \hat{a}_{\mathbf k_2} \hat{a}_{\mathbf k_1} \rangle  $. After applying a generalization of Wick's theorem to locally equilibrated systems \cite{gykw,sm}, the complete two-particle distribution  in such cases can be written as 
$N_2(\mathbf k_1,\mathbf k_2)\!=\!\omega_{\mathbf k_1} \omega_{\mathbf k_2} \, \langle
\hat{a}^\dagger_{\mathbf k_1} \hat{a}^\dagger_{\mathbf k_2} \hat{a}_{\mathbf k_2} \hat{a}_{\mathbf k_1} \rangle = 
\!\omega_{\mathbf k_1} \omega_{\mathbf k_2}  
\Bigl[\langle \hat{a}^\dagger_{\mathbf k_1} \hat{a}_{\mathbf k_1}\rangle \langle \hat{a}^\dagger_{\mathbf k_2} \hat{a}_{\mathbf k_2} \rangle + \langle \hat{a}^\dagger_{\mathbf k_1} \hat{a}_{\mathbf k_2}\rangle
\langle \hat{a}^\dagger_{\mathbf k_2} \hat{a}_{\mathbf k_1} \rangle + \langle
\hat{a}^\dagger_{\mathbf k_1} \hat{a}^\dagger_{\mathbf k_2} \rangle \langle \hat{a}_{\mathbf k_2}
\hat{a}_{\mathbf k_1} \rangle\Bigr]\!.   
$ 

For estimating the above expectation values the asymptotic operators $(\hat{a}, \hat{a}^\dagger)$ are first written in terms of the ones in-medium, $(\hat{b}, \hat{b}^\dagger)$. These last two opeartors diagonalize the full, in-medium Hamiltonian, i.e.,  $\hat{H}_m = \hat{H}_0 + \hat{H}_1 = \int d^3 k \; \Omega_k \hat{b}_k^\dagger \hat{b}_k$; the asymptotic Hamiltonian is $\hat{H}_0 = \int d^3 k \; \omega_k \hat{a}_k^\dagger \hat{a}_k$ and $\hat{H}_1$ is proportional do the mass-shift. These thermal averages are calculated by means of the density matrix operator, $\hat{\rho}$, as $\langle \hat{O} \rangle = Tr(\hat{\rho} \hat{O})$, where 
$\hat{\rho} = \frac{1}{Z} \exp \big( -\frac{1}{T} \frac{V}{2\pi^3} \int d^3 k \; \Omega_k \hat{b}_k^\dagger \hat{b}_k \big)$, $Z=Tr(\hat{\rho})$ and $T$ is the temperature. 
The resulting correlation function, given by the ratio in Eq. (\ref{corrgen}), can be written as

{\small

\be C_2({\mathbf k_1},{\mathbf k_2})=1+\frac{|G_c({\mathbf k_1},{\mathbf k_2})|^2}{G_c({\mathbf k_1},{\mathbf k_1})
G_c({\mathbf k_2},{\mathbf k_2})} + \frac{|G_s({\mathbf k_1},{\mathbf k_2})
|^2}{G_c({\mathbf k_1},{\mathbf k_1}) G_c({\mathbf k_2},{\mathbf k_2})}, 
\label{fullcorr} \ee 
}
where the denominators represent the product of the two spectral distributions, 
$G_c({\mathbf k_i},{\mathbf k_i}) = N_1(\mathbf k_i)  = \omega_{\mathbf k_i} \frac{d^3N}{d\mathbf k_i} =  \omega_{\mathbf k_i}\, \langle a^\dagger_{\mathbf k_i} a_{\mathbf k_i} \rangle$. 
In the second term, the numerator is written in terms of the chaotic amplitude, $G_c({\mathbf k_1},{\mathbf k_2}) = \sqrt{\omega_{{\mathbf k}_1} \omega_{{\mathbf k}_2} }
\langle \hat{a}^\dagger_{{\mathbf k}_1} \hat{a}_{{\mathbf k}_2}\rangle$, and is originated in the indinstinguibility of the two identical mesons (either $\phi \phi$ or $\pi^0 \pi^0$, in the current discussion), reflecting their quantum statistics. In normal conditions the third term, whose numerator is proportional to the squeezed amplitude,  $G_s({\mathbf k_1},{\mathbf k_2}) = 
\sqrt{\omega_{{\mathbf k}_1} \omega_{{\mathbf k}_2} }\langle \hat{a}_{
{\mathbf k}_1} \hat{a}_{{\mathbf k}_2} \rangle $, does not contribute. However, if the interactions in the hot and dense medium lead to mass-modification, the third term triggers this striking particle-antiparticle correlation. The three terms in Eq. (\ref{fullcorr}) contribute together in the case of neutral bosons that are their own antiparticles, such as $\phi \phi$ or $\pi^0 \pi^0$. For charged bosons , such as $\pi^\pm$ or $K^\pm$, these terms split in two separate correlation functions, envolving different pairs of particles. Thus, the first two terms in Eq.(\ref{fullcorr}) correspond to the HBT correlation, $C_c({\mathbf k_1},{\mathbf k_2})=1+\frac{|G_c({\mathbf k_1},{\mathbf k_2})|^2}{G_c({\mathbf k_1},{\mathbf k_1}) G_c({\mathbf k_2},{\mathbf k_2})}$, for identical particle pairs ($\pi^\pm \pi^\pm$ or $K^\pm K^\pm$). On the other hand,  the sum of the first and the last terms leads to  the squeezed particle-antiparticle correlation, $C_s({\mathbf k_1},{\mathbf k_2})=1+ \frac{|G_s({\mathbf k_1},{\mathbf k_2})|^2}{G_c({\mathbf k_1},{\mathbf k_1}) G_c({\mathbf k_2},{\mathbf k_2})}$, for particle-antiparticle pairs ($\pi^\pm \pi^\mp$ or $K^\pm K^\mp$). 


\section{ Results on squeezed correlations} 


Initial studies of the problem were performed for a static, infinite medium \cite{acg99,pkchp01}. Later, it was extended to the case of finite-size systems expanding with moderate radial flow \cite{phkpc05}. For the sake of simplicity a non-relativistic treatment with flow-independent squeezing parameter was considered there, which allowed to obtain analytical expressions for both the squeezed and the femtoscopic correlation functions \cite{phkpc05}. 
However, those studies focussed on the behavior of the maximum of the squeezed correlation function, $C_s({\mathbf k},-{\mathbf k},m_*)$, in terms of  modified mass, $m_*$, for particle-antiparticle pairs with exactly back-to-back momenta, ${\mathbf k_1}\!\!=\!\!-{\mathbf k_2}\!\!=\!\!{\mathbf k}$ \cite{phkpc05,pscn08}. In studies of the HBT effect, this investigation would correspond to focusing on the behavior of the $\lambda$-parameter, i.e., the intercept of the correlation function,  for identical particles with exactly identical momenta. 

Although important for theoretically understanding the finite size and flow effects on the squeezed correlation function, the systematic study of $C_s({\mathbf k},-{\mathbf k},m_*)$ does not represent a practical tool to look for the BBC's experimentally. In reality, the momenta of the two detected particle are never exactly back-to-back and the in-medium shift in the hadronic mass is not a quantity measurable in the detector. 
For an empirical search of the BBC signal, and 
considering the non-relativistic context of  Ref. \cite{phkpc05}, we suggest the following. First, select the particle and the antiparticle from the same event, with momenta $(\mathbf k_1,\mathbf k_2)$, and 
combine them to form the pair average and relative momenta, respectively, $\mathbf K\!=\!\frac{1}{2}( \mathbf k_1+\mathbf k_2)$ and  $\mathbf q\!=( \mathbf k_1-\mathbf k_2)$.  
 Then, analyze the squeezed correlation function in terms of these variables, similarly to what is done in HBT. 
 The maximual value of the BBC effect is reached for exactly back-to-back pairs, ${\mathbf k_1}\!=\!-\!{\mathbf k_2}\!=\!{\mathbf k}$, being 
located around $\mathbf{K_{_{12}}} \approx 0$. 
Therefore, the squeezed correlation function should then be investigated by varying $\mathbf K_{12}$ in the region where it is small, for several 
values of $\mathbf q_{12}$, i.e., $C_s({\mathbf k_1},{\mathbf k_2}) \rightarrow 
C_s({\mathbf 2 K_{12}},{\mathbf q_{12}})$. 

The squeezed correlation function for $\phi \phi$ pairs is obtained by inserting, in Eq.(\ref{fullcorr}), the squeezing amplitude and the spectra extracted from the results in Ref.\cite{phkpc05}, and rewritten in terms ${\mathbf K_{12}}$ and  ${\mathbf q_{12}}$, respectively, as  
\begin{widetext}
\be 
G_s(\mathbf{k}_1,\mathbf{k}_2)=\frac{E_{_{1,2}}c_{_0}s_{_0}}{(2\pi)^\frac{3}{2}} \Bigl\{ R^3 e^{-2 R^2 \mathbf{K}_{12}^2}
+ 2n^*_0 R_*^3 e^{-2 R_*^2 \mathbf{K}_{12}^2}
 \exp{\Bigl[-\frac{\mathbf{q}\;^2_{12}}{8m_* T}\Bigr]}  \exp{\Bigl[-\frac{\mathbf{K}_{12}^2}{2 m_* T_*}\Bigr]} \exp\Bigl[-\frac{i m\langle u\rangle R}{2m_* T_*}  (2 \mathbf{K}_{12})^2 \Bigr] \Bigr\}\!, 
\;\label{squeezampl}\ee 
\be 
\!\!\!G_c(\mathbf{k}_i,\mathbf{k}_i) =\frac{E_{i,i}}{(2\pi)^\frac{3}{2}} \Bigl\{|s_{_0}|^2R^3 + 2n^*_0 R_*^3  ( |c_{_0}|^2 +  |s_{_0)}|^2)  
\exp\Bigl[-\frac{(\mathbf{K}_{12} \pm \frac{1}{2}\mathbf{q}_{12})^2}{2m_* T}\Bigr]  \Bigr\},  \label{spectrum}\ee 
\end{widetext} 
where $\mathbf{K}_{12} \pm \frac{1}{2}\mathbf{q}_{12}=\mathbf{k}_{i}$, with $ i=1,2$ are the individual momenta, and considering the region where the contribution of the middle term (HBT part) in Eq. (\ref{fullcorr}) is negligible. 
The flow-modified radius and temperature in Eq.(\ref{squeezampl}) and (\ref{spectrum}),  are given, respectively, as $R_*=R\sqrt{T/T_*}$ and $T_*=T+\frac{m^2\langle u\rangle^2}{m_*}$, as in Refs. \cite{phkpc05,pscn08}. 
We adopt natural units, $\hbar = c =1$, througout this article. 

Extending the analogy with usual procedures adopted in HBT, we could think that background could be chosen experimentally by combining particle-antiparticle pairs from different events. This choice corresponds to consider an uncorrelated pair, free from identical particle exchange effects. However, 
this is different in the BBC case, since the squeezing factor appears in the denominator of the correlation function as well. 
Therefore, on searching for the effect of squeezing, we should consider the product of the spectra of the particle and the antiparticle in the denominator of $C_s({\mathbf 2 K_{12}},{\mathbf q_{12}})$, each one writen as in Eq. (\ref{spectrum}). 

Naturally, the analysis in terms of the variable $2\mathbf K$ would not be suited for a genuine relativistic treatment. In this case, a relativistic four-momentum variable can be constructed as $Q^\mu_{back}=(\omega_1-\omega_2,\mathbf k_1 + \mathbf k_2)=(q^0,2\mathbf K)$, first introduced in \cite{pscn08}. In fact, it is preferable to redefine this variable  as $Q^2_{bbc} = -(Q_{back})^2=4(\omega_1\omega_2-K^\mu K_\mu )$, since its  non-relativistic limit is $Q^2_{bbc} \rightarrow (2 \mathbf K)^2$, recovering the average momentum of the pair introduced above. 

In Ref.\cite{pscn08}, we presented some introductory results on 
$\phi\phi$ squeezed correlations in terms of $|\mathbf{K_{_{12}}}|$ and $|\mathbf{q_{_{12}}}|$, but most of the plots shown there  were obtained by attributing precise values to the variables in Eq. (\ref{fullcorr}), (\ref{squeezampl}), and(\ref{spectrum}). In the current analysis, however, we follow the above procedure in a more realistic estimate, in which the momenta are generated in a simulation and 
 then combined to form $|\mathbf{K_{_{12}}}|$ and $|\mathbf{q_{_{12}}}|$. The binning in such a simulation should reflect the finite experimental resolution in momentum, and experimental acceptance cuts could also be introduced, whenever available. It should be stressed that the choice of $\phi$ mesons considered here was made as a means to illustrate the proposed method, for two main reasons: $\phi$'s are their own antiparticles and their large mass validates the non-relativistic approximation considered here.  

\begin{figure}[htb]
\begin{center}
 \includegraphics[height=.24\textheight]{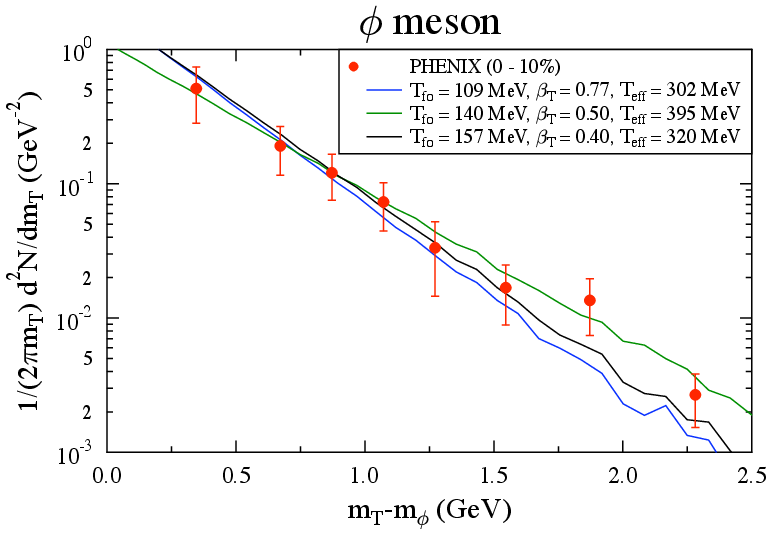}
\caption{(Color online) Generated transverse mass distributions (arbitrary normalization) are compared with PHENIX data for three sets of fit parameters: i) the freeze-out temperature $T_{fo}=109$ MeV and flow velocity $\beta_T=0.77$, ii) $T_{fo}=157$ MeV and $\beta_T=0.4$, used in PHENIX simulation, and 
 $T_{fo}=140$ MeV and $\beta_T=<u>=0.5$, 
 adopted here.} \label{spectrumphi}
\end{center}
\end{figure}

In the simulation we generated the momenta of the $\phi$ mesons with the following prescription. For roughly mimicking the experimental cuts, based on PHENIX data on $\phi$'s \cite{phenixphi}, we introduced some simple geometrical selection criteria in their generation. 
This was done by merely considering the cuts in azimuthal angle and in the pseudo-rapidity, as well as by selecting the momentum region. 
In Fig. \ref{spectrumphi}, we show the transverse momenta generated in the simulation, as compared to experimental data points, for three sets of parameters corresponding to the temperature, $T$, and to the radial flow velocity, $\langle u \rangle$. 
All three sets are in reasonable agreement with the measured distributions. In the remainder of this 
work, we fix $T=140$ MeV. 

\begin{figure*}[htb] 
\begin{center}
\includegraphics[height=.25\textheight]{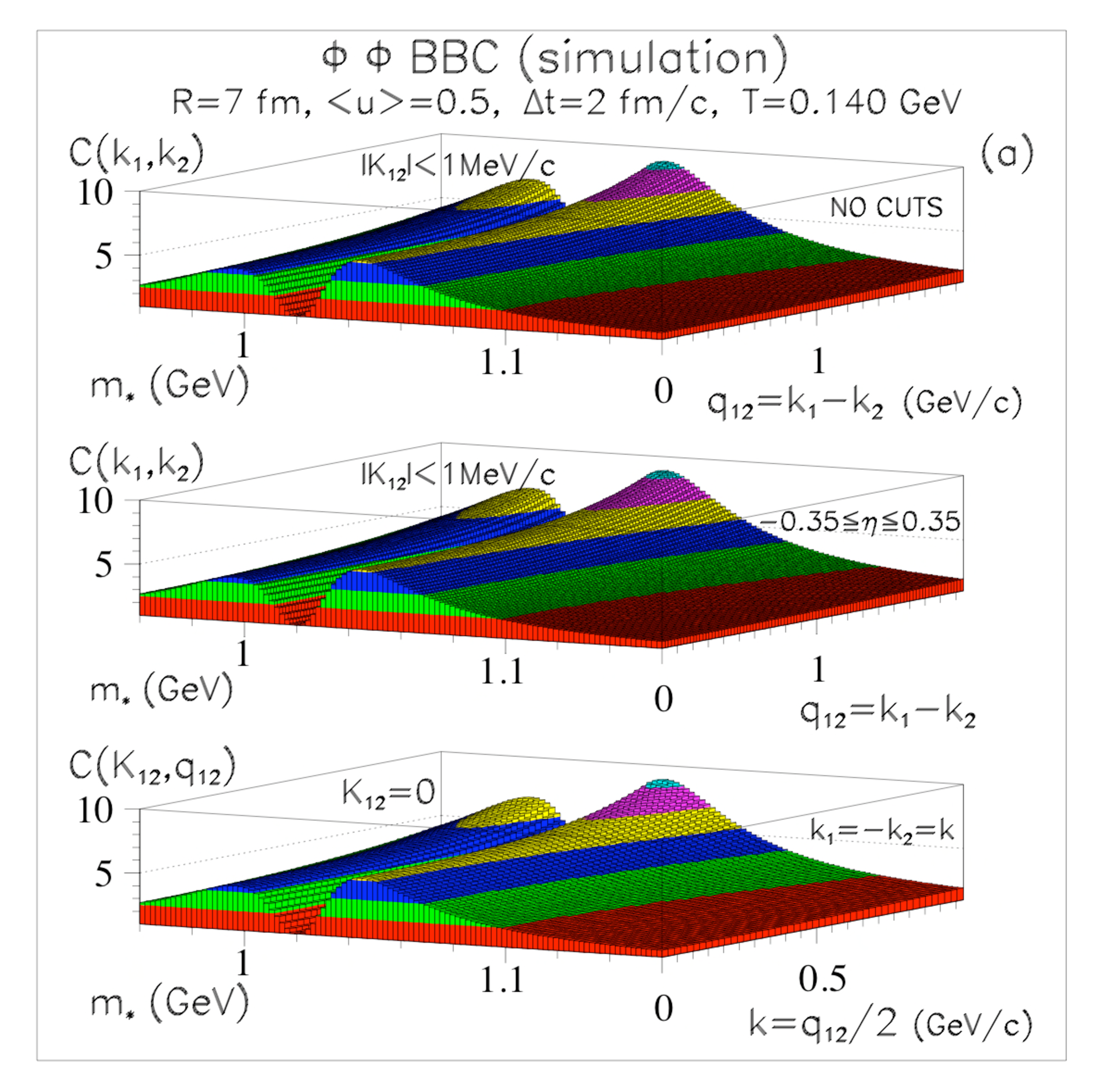}
\includegraphics[height=.25\textheight]{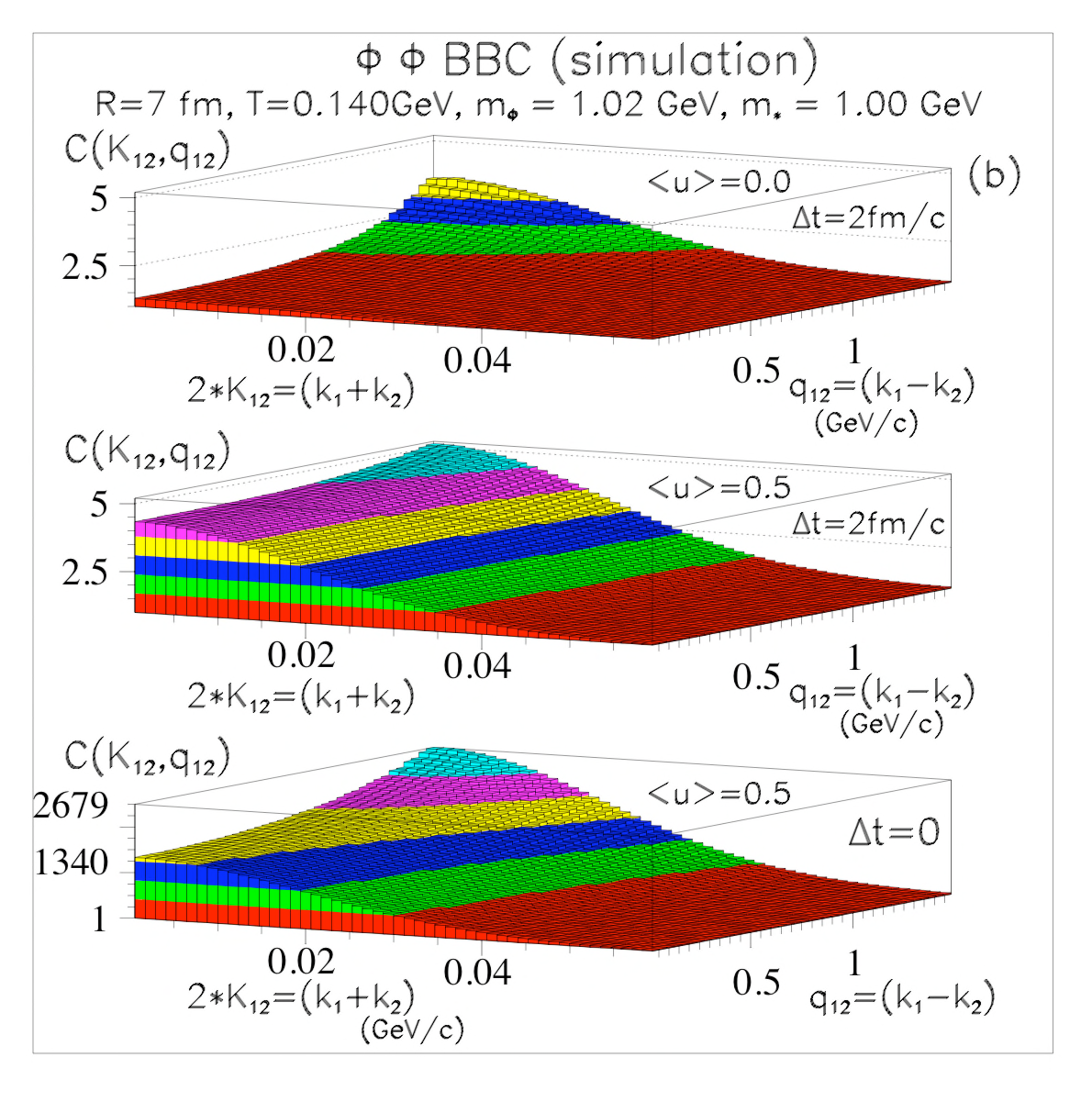}
\includegraphics[height=.245\textheight]{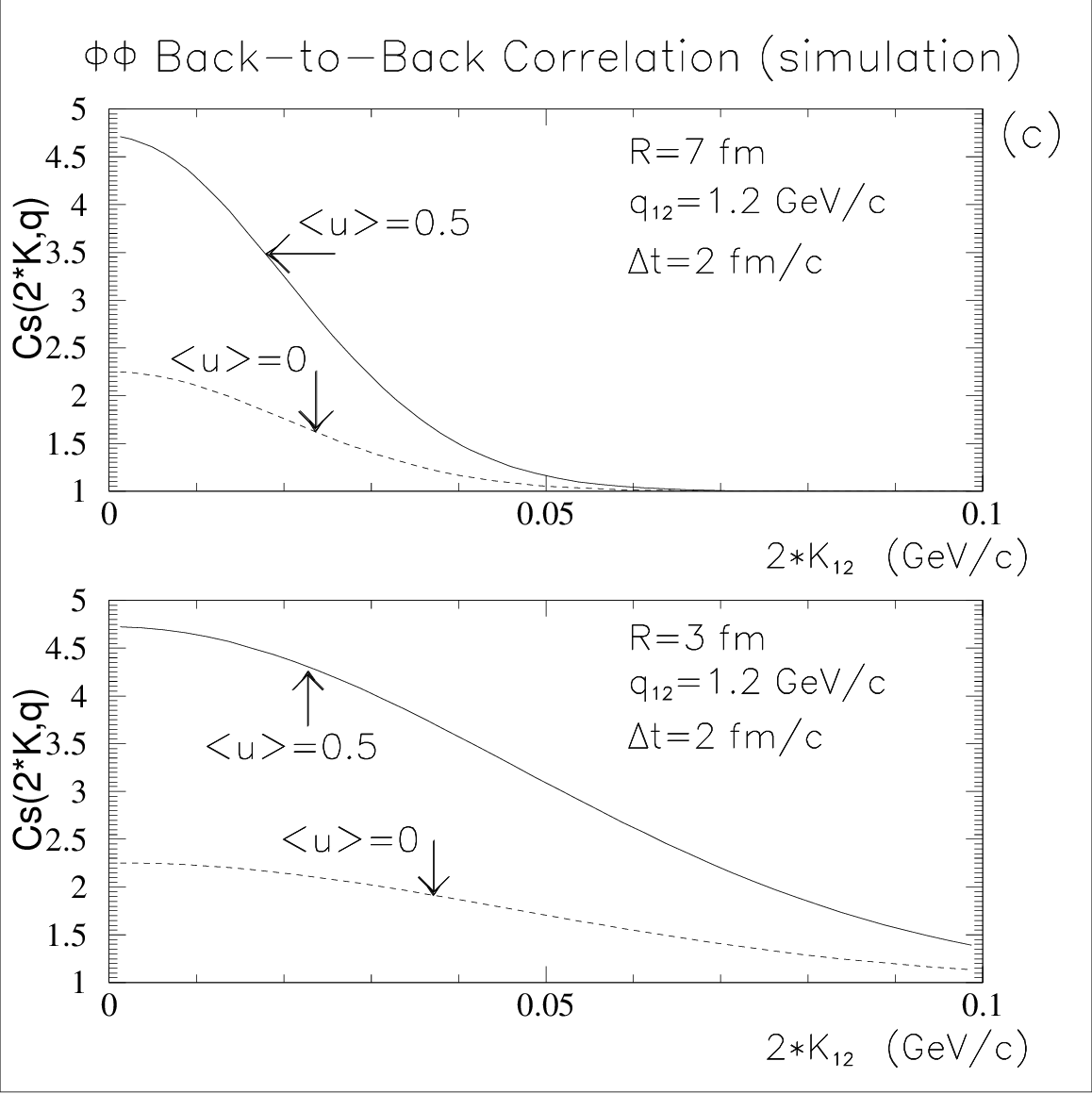}
\caption{(Color online) Part (a) shows the squeezed  
correlation function in terms of the shifted mass, $m_*$, and the pair relative momentum, $|\mathbf q_{_{12}}|$.  Results from simulation (top two plots) are compared with those from calculation, with fixed $K_{12}=0$ (bottom). Part (b) shows the advertised plots for the squeezed correlation function to be searched for experimentally, with $m_*=1$GeV. The striking effect of finite emission times is shown by comparing the middle and the bottom plots in part (b). 
Part (c) shows the inverse width of the BBC function reflecting the radius of the squeezing region, for $R=7$ fm (top) and $R=3$ fm (bottom).}\label{squeezcorr}  
\end{center}
\end{figure*} 

We then combine the momenta of the $\phi\phi$ particle-antiparticle pairs to calculate the squeezed correlation function, as explained above. 
The top two plots correspond to the simulated $C_s(m_*, \mathbf{q_{_{12}}})$, keeping the average momentum of the pair in a small interval, $|\mathbf{K_{_{12}}}| \le1$ MeV. In the top most plot no cuts were considered, but in the middle plot of Fig. \ref{squeezcorr} (a),  a rough version of the cuts from Ref.\cite{phenixphi} were introduced in the simulation. No sensitivity to those cuts is apparent. At last, 
for cross-checking the simulation code, we compare the squeezed correlation functions estimated with the generated pairs, with that  obtained by considering pairs with exactly back-to-back momenta. This is seen 
in the plot at the bottom of Fig. \ref{squeezcorr}(a), showing the histogram obtained by attributing exact values to $\mathbf{q_{_{12}}}$, and also fixing  $\mathbf{K_{_{12}}}\equiv 0$,  similarly to what was shown in Ref. \cite{pkchp01,phkpc05,pscn08}. This 
plot in the bottom shows close similarity with the first two, as would be required. 

Since the strength of the squeezed correlation is expected to be significant in the low-$|\mathbf{K_{_{12}}}|$
region, we also suggest to plot  $C_s(\mathbf{K_{_{12}}}, \mathbf{q_{_{12}}})$ as a function $2 \mathbf{K_{_{12}}}$, for 
enlarging the average momentum region where the correlation intensity can be significantly above unity. 
Fig. \ref{squeezcorr} (b) shows the result for the squeezed correlation function from simulation, 
obtained considering the static case ($\langle u \rangle=0$), on top, and the case with radial flow ($\langle u \rangle=0.5$), in the middle. In both, we considered that the particles were emitted in a finite interval, $\Delta t = 2$ fm/c, during which the emission decreases 
due to a Lorentzian distribution in time, {\small $|F_s(\Delta t)|^2=[1+(\omega_1+\omega_2)^2\Delta t ^2]^{-1}$} \cite{acg99,pkchp01,phkpc05,pscn08}, which multiplies the third term in Eq. (\ref{fullcorr}). This factor reduces the signal by almost three orders of magnitude, as compared to a instantaneous emission  
($\Delta t =0$). This can be seen by comparing the plot in the middle of Fig. \ref{squeezcorr}(b), with the one in the bottom. 
We note that, although the reduction of the strength caused by a finite emission period is dramatic, the intensity of the $\phi \phi$ BBC correlation is still sizable, suggesting that its experimental search is indeed promising. Another particular emission time distribution will be discussed below. 


From Figs. \ref{squeezcorr}(b) we see that, in the absence of flow, the squeezed correlation 
grows faster from smaller to higher values $|\mathbf q_{_{12}}|$ than in the presence of flow. However, this last one is stronger even in the low $|\mathbf q_{_{12}}|$ region, showing that the presence of flow could enhance the signal's intensity 
over a wider region of the ($|\mathbf K_{_{12}}|$,$|\mathbf q_{_{12}}|$)-plane. The size of the squeezing region is reflected in the inverse width of the curves as a function of $2 |\mathbf K_{_{12}}|$, being narrower (broader) for larger (smaller) radii. In Figs. \ref{squeezcorr}(c), we compare the squeezed correlation functions considering that the system radius is $R=7$ fm, also used in all the above calculation, to the case of a system with smaller extent of the squeezing region, with $R=3$ fm.

 Returning to the discussion about the time emission distribution, we 
should emphasize that we do not know a priori how this emission should proceed, mainly in the simple model adopted here. On the other hand, the lack of knowledge of such distribution, naturally does not invalidate the experimental search of the hadronic squeezed correlations. The squeezing is a fundamental phenomenon already detected in quantum optics and it should be empirically observed in relativistic heavy ion collisions, if the hadron masses are modified in-medium by some mechanism. Therefore, as a common practice in Physics, we should look for the effect experimentally and, once it is discovered, we could try to explain its time emission process by means of a suitable model. Nevertheless, motivated by an analysis made by the PHENIX Collaboration\cite{levy-phenix} we investigated in Ref. \cite{danuce-M,danuce} a different emission distribution in time, by considering the effects of a a symmetric, $\alpha$-stable L\'evy distribution, i.e., 
$
|F(\Delta t)|^2=\exp\{-[\Delta t (\omega_1+\omega_2)]^\alpha\},  \label{levy}
$
on the squeezed correlation function of $K^+ K^-$ pairs. 
This functional form had been fitted to two- and three-particle Bose-Einstein correlation functions. The values fitted to date different values of the distribution index, $\alpha=1.0$ or $\alpha=1.35$, were fitted to data, 
depending on the region investigated of the particles' transverse momentum or transverse mass. 
Briefly summarizing those effects, we concluded that, for $\alpha=1$,  such a time factor acting on the squeezing correlation function reduces its intensity even more dramatically than the Lorentzian factor discussed here. However, it would still lead to measurable quantities, mainly if the emission lasted a short period of time, of about $\Delta t=1$ fm/c. However, if Nature favors the higher value, $\alpha=1.35$,  this would result in a very small deviation from unity, not detectable by the method proposed here. Naturally, in the case of $\phi$ pairs, even for $\alpha=1$ and $\Delta t=1$ fm/c the intensity would be negligibly small for a L\'evy-type distribution, due to their large asymptotic mass. In spite of that, as discussed above, we do not know a priori the preferred form chosen by Nature for the emission process, which in itself does not invalidate the experimental search of this phenomenon. For this reason and for continuing the illustration of the proposed method to search for the hadronic squeezed states, in the remainder of this paper we attain our discussion to the Lorentzian time distribution, comparing it to the instantaneous emission process only.

\section{Effects of squeezing on $\phi \phi $ HBT correlations} 

We next discuss how the HBT correlation function could be affected by the in-medium mass-shift. 
Contradicting early expectations that the thermalization would wash out any trace of mass-shift in this type of correlation, we find its in-print in HBT, reflecting the presence of the squeezing factor, $f_{i,j}(m,m_*)$, in the chaotic amplitude. This can be inferred from the analytical results in previous papers\cite{acg99,pkchp01,phkpc05}. However, the strength of the squeezing effect on the two-identical particle correlation was not carefully investigated in those references. 

The analytical form of the HBT correlation function is obtained by substituting the chaotic amplitude\cite{phkpc05}, 
\begin{widetext}
\be 
G_c(\mathbf{k}_1,\mathbf{k}_2) = \frac{E_{_{1,2}}}{(2\pi)^\frac{3}{2}} \Bigl\{ |s_{_0}|^2 R^3 e^{-\frac{1}{2} R^2 \mathbf{q^2_{12}}}\!\!+ 
n^*_0 R_*^3 (|c_{_0}|^2+|s_{_0}|^2) 
e^{-\frac{1}{2}R^2_* \mathbf{q_{12}^2}} \; e^{-\frac{\mathbf{K_{12}^2}}{2 m_* T_*}} \; e^{-\frac{\mathbf{q_{12}^2}}{8m_* T}} \exp\Bigl[
\frac{im\langle u\rangle R}{m_* T_*}
\mathbf{K}_{12} . \mathbf{q}_{12}\Bigr]  \Bigr\}, \label{chaotampl} \ee 
\end{widetext}
together with the spectrum given in Eq. (\ref{spectrum}), into Eq.(\ref{fullcorr}). The finite emission time factor, multiplying the square modulus of Eq. (\ref{chaotampl}), is now {\small $F_c(\Delta t)=[1+(\omega_1-\omega_2)^2\Delta t ^2]^{-1}$}. For stressing the HBT effects in the $\phi \phi$ case, we selected the 
region (small $|\mathbf{q_{_{12}}}|$) where the particle-antiparticle correlation is not significant and the HBT is relevant to Eq.(\ref{fullcorr}). 

\begin{figure}[htb]
\begin{center}
\includegraphics[height=.255\textheight]{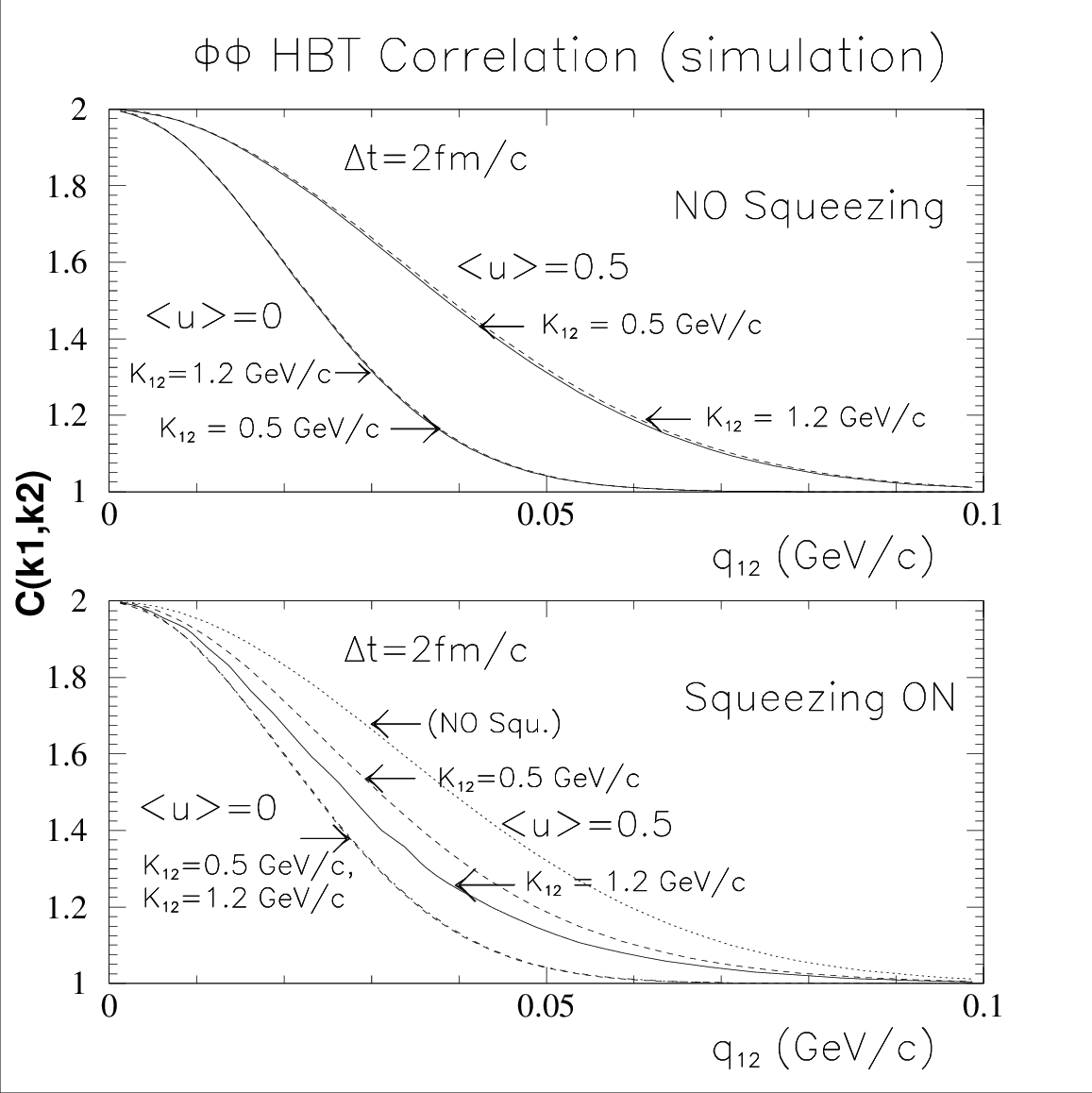}
\caption{The plots show the HBT correlation function in the absence of squeezing (top) and when it is present (bottom). }\label{hbt-on}
\end{center}
\end{figure}

This is seen in Fig. \ref{hbt-on}. 
The top part shows the effect of radial flow alone on the HBT correlation function, while in the bottom, the joint effects of flow and squeezing are shown. We see that, without squeezing, the flow broadens the correlation curves, as expected, since the 
expansion reduces the size of the region accessible to interferometry. 
When the squeezing effects are present, they seem to oppose to the flow effects, almost canceling the broadening of the 
correlation function due to flow for large $|\mathbf K|$,   another striking indication of in-medium mass modification.

\section{Conclusions}
In this work 
we suggest an effective way to search for the squeezed bosonic 
correlations in heavy ion collisions at RHIC and, soon, at the LHC.
We argue that the suitable variable to experimentally search for the squeezed correlation function is the average momentum of the pair, $2 |{\mathbf K_{12}}|$, the non-relativistic limit of the  relativistic variable, $Q_{bbc}=2\sqrt{(\omega_1\omega_2-K^\mu K_\mu )}$ \cite{pscn08}. 
We show that, in the presence of flow, the signal is expected to be stronger over the momentum regions shown in the plots, i.e., roughly for $0 \lesssim |2 \mathbf K| \lesssim 100$ MeV/c (depending on $R$) and $500 \lesssim |\mathbf q| \lesssim 1500-2000$ MeV/c, suggesting that flow may enhance the strength of the BBC signal, facilitating its experimental discovery. 
Another important result found within our 
simple non-relativistic model is that the squeezing could also distort the HBT correlation function, leading to effects opposing 
those of flow, almost neutralizing it for large values of $|\mathbf K_{_{12}}|$. 
For emphasizing the dramatic effects induced by in-medium hadronic mass modification on the correlation functions, we chose a constant mass-shift that leads to the maximal intensity, 
based on results of Fig. (\ref{squeezcorr}). For $\phi \phi$ mesons, this corresponds to $m*\approx 1 $ GeV, roughly a $2\%$ reduction in the $\phi$ mass, as compared to its asymptotic mass ($m=1.02$ GeV). As 
stressed before, a more realistic treatment 
should consider a detailed prescription for the mass modification, based on models that predict its dependence on the particles' momenta and its distribution in the hot and dense system. 

The above procedure is also applicable to other particles, such as kaons. 
The corresponding results \cite{danuce-M} are discussed in Ref. \cite{danuce}.
 Finally, it is important to note that all the effects 
 shown here should exist only if the particles 
 have 
 their mass modified 
 in the hot and dense medium.  If no modification happens, the squeezed correlation functions would be flat unity, and the HBT correlation functions would behave as usual. However, if the particles' masses are indeed modified, the experimental discovery of squeezed particle-antiparticle correlation (and the distortions pointed out in the HBT correlations) would be an unequivocal signature of in-medium modifications by means of hadronic probes!  


\section{Acknowledgements} 
We are grateful to T. Cs\"org\H{o} and M. Nagy, who proposed the $Q^\mu_{back}$ variable, for fruitful discussions and  
comments on the text. We are also thankful to H. Takai 
for a careful reading and suggestions on the manuscript. OSJ acknowledges funding from CNPQ.
 

\end{document}